\documentclass[useAMS]{gGAF2e}

\begin{document}
\doi{-}
 \issn{-} \issnp{-} \jvol{-} \jnum{-} \jyear{-} 

\markboth{K.N. Gourgouliatos AND N. Vlahakis}
\\
\title{Relativistic expansion of a magnetized fluid}

\author{K.N. GOURGOULIATOS${\dag}$$^{\ast}$\thanks{$^\ast$Corresponding author. Email:kgourgou@purdue.edu 
\vspace{6pt}} and N. VLAHAKIS${\ddag}$\\\vspace{6pt}  ${\dag}$ University of Cambridge, Institute of Astronomy, 
CB3 0HA, UK\\ 
${\dag}$Purdue University, Department of Physics, 525 Northwestern Avenue, W. Lafayette, IN 47906, USA \\ ${\ddag}$Section of Astrophysics, Astronomy and Mechanics, Physics Department, University of Athens, 15784 Zografos, Athens, Greece\\\vspace{6pt}}

\maketitle

\begin{abstract}
We study semi-analytical time-dependent solutions of the relativistic magnetohydrodynamic (MHD) equations for the fields and the fluid emerging from a spherical source. We assume uniform expansion of the field and the fluid and a polytropic relation between the density and the pressure of the fluid. The expansion velocity is small near the base but approaches the speed of light at the light sphere where the flux terminates. We find self-consistent solutions for the density and the magnetic flux. The details of the solution depend on the ratio of the toroidal and the poloidal magnetic field, the ratio of the energy carried by the fluid and the electromagnetic field and the maximum velocity it reaches.  
\end{abstract}
\bigskip

\bigskip

\section{Introduction}
\label{introduction}

\cite{P2005} presented a study of time dependent, relativistic, force-free, ideal MHD in the absence of matter by imposing a self-similar form for the solutions of the problem. In his pioneering work he assumed that time-dependence appears only through a dimensionless variable which contained in addition to the time, the speed of light and the radial distance from the centre. Then, he found a relativistic form of the Grad-Shafranov differential equation \citep{GR1958, S1958, S1966}. However, he did not take into account any pressure due to the surrounding plasma. In this paper we built up on Prendergast's work by taking into account the effect of pressure while still aiming for equilibrium solutions. The presence of pressure leads to extra forces and the electromagnetic field is no more force-free, but now it is the net force due to the pressure and the electromagnetic field that has to be zero, thus the electromagnetic field interacts with the plasma. This interaction leads, in the most general case, to a set of non-linear partial differential equations. In this work we study forms of these equations permitting analytical solutions, so that we can have a general picture of such systems. 

Apart from Prendergast, problems of relativistic MHD have been studied by other authors. \cite{CLB1991} studied the asymptotic behaviour of steady, fully relativistic, axisymmetric, hydromagnetic winds and found that the flux surfaces take the form of cylinders and parabolas around the rotation axis, \cite{CLB1992} studied self-similar solutions for relativistic winds driven by rotating magnetic fields. \cite{C1994,C1995} studied the full ideal MHD problem for steady state cold outflows and he found that the form of the solution depends on the amount and the distribution of the electric current, he also presented self-similar solutions for the same problem. \cite{F1997} and \cite{FG2001} studied force-free magnetospheres near rotating black holes and found strong evidence for a hollow jet structure and applied these solutions to galactic superluminar sources. \cite{D2005} simulated the problem of magnetic accretion in a rotating black hole taking into account the Kerr metric. \cite{HN2003} found stationary solutions for axisymmetric, polytropic, unconfined, ideal MHD wind using the WKB method. The general motivating for these studies is relativistic outflows in the form of jets and winds related either to AGN or to stellar mass black holes. They are mainly numerical and as such they have limitations on the parameters chosen and also on the trial functions employed to solve the systems of the partial differential equations. Our treatment leads to an analytical solution, where the system of partial differential equations simplifies by the use of self-similarity and finally we only solve an ordinary differential equation numerically. Analytical solutions allow an easier study of the parameter space and a better insight on the physical behaviour, however, they require simplifications and special boundary conditions. 

A way to introduce time dependence in the problem is by imposing a temporally self-similar solution. This type of self-similarity leads to solutions which are functions of a new variable $\tilde{x}=r^{\lambda}v(r,t)$, which is a product of a power of the spatial coordinate and a combination of time and the spatial coordinate. This method has been used in similar problems, of which the best known is the blast wave solution by \cite{S1946} and \cite{T1950}. In this case a relation between the expansion radius and time is found by dimensional analysis of the physical quantities involved in the system. Then the equations are solved and provide the details of the explosion. Examples of this type of self-similarity in force-free relativistic MHD can be found in \cite{P2005,GL2008,G2009}.

The other form of self-similarity we are going to use is related to the separation of variables. In problems depending on two spatial variables, one can seek solutions which are products of a function of the angular coordinate and a function which depends on the distance from the origin. If a suitable form is imposed for the angular function and then the equation is solved numerically for the other function we find the meridionally self-similar solutions. In the case of radially self-similar solutions a suitable form is imposed for the radial function, and then the equation is solved for the angular part of the problem. Examples of MHD problems solved by this form of self-similarity can be found in \cite{BP1982,LB1994,ST1994,VT1998}.
 
In the problem we are solving in this paper, we use the self-similarity technique in two steps. In the initial formalism of the problem we demand that the time evolution of the fields will only appear through the dimensionless combination of $v=r/(ct)$. Then, we rewrite the system of the partial differential equations using this new variable. We observe that the system separates by imposing meridionally self-similar solutions. Thus, by choosing a class of those solutions the problem reduces to the solution of an ordinary differential equation, which we integrate numerically. These numerical solutions depend on the boundary conditions and the parameters chosen. We explore the parameter space and discuss the significance of the parameters chosen and their implications for the nature of the system. \cite{TS2007} studied a similar problem of an expanding magnetized fluid in the non-relativistic limit while taking into account Newtonian gravity. 

\section{Formulation of the problem}
\label{formulation}

We consider a system containing an electromagnetic field ${\bf E}$, ${\bf B}$ as measured by an observer stationary relative to the centre of the system and a fluid of rest density $\rho_0$ and pressure $p$. The system expands uniformly with scaled velocity 
\begin{eqnarray}
{\bf v} =\frac{r}{ct}{\bf \hat{e}}_r\,.
\label{velocity}
\end{eqnarray}
As a result of the assumed uniform expansion, the Lagrangian derivate of the velocity vanishes and thus, each element of the system moves with constant velocity. This is a requirement for an equilibrium expansion. Had the Lagrangian derivative not been zero then the same fluid element would have suffered some acceleration or deceleration and the net force would have not been zero. The second assumption is that of axial symmetry, thus the physical quantities do not depend on the $\phi$ coordinate. The third assumption is the ideal MHD approximation, therefore in the frame of the fluid the electric field vanishes, thus 
\begin{eqnarray}
{\bf E}=-{\bf v}\times {\bf B}\,.
\label{ohm}
\end{eqnarray}
The electromagnetic field has to satisfy Maxwell's equations
\begin{eqnarray}
\nabla \cdot {\bf B}=0\,,
\label{divB}
\end{eqnarray}
\begin{eqnarray}
\nabla \times {\bf E}=-\frac{1}{c}\frac{\partial {\bf B}}{\partial t}\,,
\label{faraday}
\end{eqnarray}
\begin{eqnarray}
\nabla \cdot {\bf E}=\frac{4 \pi}{c}j^0\,,
\label{gauss}
\end{eqnarray}
and
\begin{eqnarray}
\nabla \times {\bf B}=\frac{1}{c}\frac{\partial {\bf E}}{\partial t}+\frac{4 \pi}{c} {\bf j}\,.
\label{ampere}
\end{eqnarray}
Equations~(\ref{gauss}) and (\ref{ampere}) which contain charge and current densities, allow us to determine these densities. The fluid has to satisfy the baryon mass conservation which is
\begin{eqnarray}
\Big(\frac{\partial}{\partial t}+c{\bf v}\cdot\nabla\Big)(\gamma \rho_0)+c\gamma \rho_0 \nabla \cdot {\bf v}=0\,,
\label{continuity}
\end{eqnarray}
where $\gamma=(1-v^2)^{-1/2}$ is the Lorentz factor and $\rho_{0}$ is the rest mass density.

The momentum equation is (see, e.g., \citealp{VK03a}) 
\begin{eqnarray}
-\gamma \rho_0\Big(\frac{\partial}{\partial t} +c{\bf v}\cdot \nabla\Big)(\xi \gamma c{\bf v})-\nabla p 
+\frac{j^0{\bf E}+{\bf j} \times {\bf B}}{c} =0\,,
\label{momentum}
\end{eqnarray}
where the relativistic specific enthalpy (over $c^2$)
for a polytrope with $\Gamma=4/3$ is 
\begin{eqnarray}
\xi=1+4\frac{p}{\rho_0c^{2}}\,.
\end{eqnarray}
We have chosen $\Gamma=4/3$ to allow self-similar solutions \citep{L1982}.
Finally the entropy equation is
\begin{eqnarray}
\Big(\frac{\partial}{\partial t}+c{\bf v}\cdot\nabla\Big)\left(\frac{p}{\rho_0^{4/3}}\right) =0\,.
\label{entropy}
\end{eqnarray}

The above system of equations has to be solved in order to determine the density and the fields. 

We express the magnetic field in terms of two quantities, $P$ and $T$. The flux function $P$ depends on $v$ and $\theta$, and is the magnetic flux that passes through a cap of semi-opening angle $\theta$ and lies in distance $v$ from the origin in the velocity space. The function $T$ is related to the toroidal component of the magnetic field. The expression of the magnetic field that by construction satisfies~(\ref{divB}) is
\begin{eqnarray}
{\bf B}=\frac{1}{2\pi r^{2} \sin\theta } \Big(\frac{\partial P}{\partial \theta}{\bf \hat{e}}_r-v\frac{\partial P}{\partial v}{{\bf \hat{e}_\theta}}+T{{\bf \hat{e}_\phi}} \Big)\,.
\label{magnetic}
\end{eqnarray}
Equation~(\ref{ohm}) gives the electric field
\begin{eqnarray}
{\bf E}=\frac{1}{2 \pi r^{2}\sin\theta}\Big(vT{{\bf \hat{e}_\theta}}+v^{2}\frac{\partial P}{\partial v}{{\bf \hat{e}_\phi}}\Big).
\label{electric}
\end{eqnarray}
The velocity of the field lines is ${\bf v}_{F}=c{\bf E} \times {\bf B}/|{\bf B}^{2}|$ and it has a $\phi$ component, this component is a geometrical effect of the expansion and the toroidal component of the magnetic field and there is no rotation of the central dipole.

For axially symmetric radial flows, the ${{\bf \hat{e}_\phi}}$ component of the momentum equation~(\ref{momentum}) yields that $\left(j^0 {\bf E}+{\bf j}\times{\bf B}\right)\cdot {{\bf \hat{e}_\phi}} = 0$, which leads to a differential equation for $P$ and $T$, 
\begin{eqnarray}
\frac{\partial T}{\partial v} \frac{\partial P}{\partial \theta} - \frac{\partial T}{\partial \theta} \frac{\partial P}{\partial v}-\frac{v^{2}+1}{v(1-v^2)}T\frac{\partial P}{\partial \theta}=0\,,
\label{Teq0}
\end{eqnarray}
or, by multiplying~(\ref{Teq0}) with $(1-v^{2})/v$,
\begin{eqnarray}
\frac{\partial}{\partial v}\Big(\frac{1-v^{2}}{v}T\Big)\frac{\partial P}{\partial \theta}-\frac{\partial}{\partial \theta}\Big(\frac{1-v^{2}}{v}T\Big)\frac{\partial P}{\partial v}=0\,.
\label{Jacobian}
\end{eqnarray}
This is the Jacobian of $P$ and $[(1-v^{2})/v]T$ with respect to $v$ and $\theta$, thus
\begin{eqnarray}
T=\gamma^2 v \beta(P)\,,
\label{Teq}
\end{eqnarray}
where $\beta(P)$ is an arbitrary function of $P$.

\section{Solution}
\label{solution}

\subsection{The entropy equation}

Equation~(\ref{entropy}) yields a relation between density and pressure $p=Q\rho_0^{4/3}$, where $Q$ is a function of $v$ and $\theta$.  We can use this equation to find the density as a function of the pressure $\rho_0 = p^{3/4}/Q^{3/4}$.

\subsection{The baryon mass conservation equation}
 
Substituting the above expression of the density in~(\ref{continuity}) we find that the pressure has a form that can be conveniently written as
\begin{eqnarray}
p= p_0 \frac{\gamma^4 v^4 }{r^4}\,,
\label{pressure}
\end{eqnarray}
where $p_0$ is a function of $v$ and $\theta$.

Then, the density is given by
\begin{eqnarray}
\rho_0= \left(\frac{p_0}{Q}\right)^{3/4} \frac{\gamma^3 v^3}{r^3} \,.
\label{density}
\end{eqnarray}

\subsection{Maxwell's equations}

By construction, the form of the magnetic field chosen satisfies~(\ref{divB}). The induction equation~(\ref{faraday}) is also satisfied for the adopted forms of the magnetic and electric fields. The other two equations have the current and charge densities that are not determined yet. By solving equations~(\ref{gauss}) and (\ref{ampere}) for $j^{0}$ and ${\bf j}$ respectively, we express these quantities in terms of the functions $P$ and $\beta$. The resulting expressions for the charge and current densities are
\begin{eqnarray}
\frac{j^{0}}{c}&=&\frac{\gamma^2 v^{2}}{8 \pi^{2}r^3}\frac{{\rm d}\beta}{{\rm d}P}\frac{\partial P}{\partial \theta}\,,
\label{charge}
\\
{\bf j}&=&\frac{c}{8 \pi^{2} r^{3} \sin \theta}\Big\{v\frac{{\rm d}\beta}{{\rm d}P}\Big[\gamma^2\frac{\partial P}{\partial \theta}{\bf \hat{e}}_r-v\frac{\partial P}{\partial v}{{\bf \hat{e}_\theta}}\Big]+ \Big[-\frac{v^{2}}{\gamma^2}\frac{\partial^{2}P}{\partial v^{2}}+2v^{3}\frac{\partial P}{\partial v}-\sin^2 \theta \frac{\partial^2 P}{\partial \left(\cos\theta\right)^2}\Big]{{\bf \hat{e}_\phi}}\Big\} \,.
\label{current}
\end{eqnarray}
Next we use these results to solve the momentum equation.

\subsection{The momentum equation}
\label{mom_section}

The momentum equation~(\ref{momentum}) contains three terms. The first one is the inertia term ${\bf f}_{\rm I}$, which is proportional to the derivative of the relativistic specific enthalpy $\xi$. The second term, ${\bf f}_{p}$, is due to the pressure gradient. The third term, ${\bf f}_{\rm em}$ is due to the electromagnetic forces. We are going to evaluate each term of this equation and seek analytical and semi-analytical solutions. The first term is
\begin{eqnarray}
{\bf f}_{\rm I}=4\gamma^2v^2\frac{p}{r} {\bf \hat{e}}_r\,.
\label{inertia_term}
\end{eqnarray}
The second term, using~(\ref{pressure}), is
\begin{eqnarray}
{\bf f}_{p}=-\frac{\gamma^4 v^4}{r^4} \nabla p_0
- 4\gamma^2v^2\frac{p}{r} {\bf \hat{e}}_r \,.
\label{pressure_term}
\end{eqnarray}
From the sum ${\bf f}_{\rm I}$ and ${\bf f}_{p}$ only the first term of ${\bf f}_{p}$ survives and the effect of inertia is cancelled by the second term of the pressure force. This is because we have chosen a configuration that expands uniformly and as such there is no acceleration on the fluid. The ${\bf f}_{\rm I}$ is a pseudo-force that appears because of the choice of the frame of reference.
 
The third  term is
\begin{eqnarray}
{\bf f}_{\rm em}=-\frac{{\cal F}}{16 \pi^3 r^4\sin^2\theta} \nabla P \,,
\label{em_term}
\end{eqnarray}
where
\begin{eqnarray}
{\cal F} =
\frac{v^2}{\gamma^2} \frac{\partial^{2} P}{\partial v^{2}}
-2v^{3}\frac{\partial P}{\partial v}
+\sin^2 \theta \frac{\partial^2 P}{\partial (\cos \theta)^2}
+\gamma^2 v^2 \beta \frac{d \beta}{dP}
\,.
\label{GS}
\end{eqnarray}
${\cal F}=0$ is the relativistic form of the Grad-Shafranov equation for uniform expansion in the force-free limit, as it was formulated by \cite{P2005}.  Indeed our study in the case of negligible pressure reduces to this equation, whose detailed study can be found in \cite{GL2008}. In the present paper we study cases where the fluid pressure is no longer negligible. The total force on a volume element due to gas pressure and electromagnetic interaction is zero. If ${\cal F} \neq 0$ the electromagnetic force is nonzero; it is normal to the magnetic field (since $\nabla P \ \bot \ {\bf B}$) and, depending on the sign of ${\cal F}$, points towards the axis or in the opposite direction.

Note that if gravity is non-negligible, a fourth term should be added on the left-hand side of the momentum equation~(\ref{momentum}) and time and space coordinates have to be modified according to the metric. This gravitational term is ${\bf f}_{G}=-\gamma^2 \rho_{0}c^{2}\xi \nabla \ln h$ (see e.g., \citealp{ML1986,Meliani2006}), where $h=(1-r_{\rm S}/r)^{1/2}$ is the redshift factor with $r_{\rm S}=2GM_{*}/c^{2}$ the Schwarzschild radius for a central mass $M_{*}$. The appearance of the redshift factor, which is solely a function of $r$, makes the separation of variables ($v\,, \theta$) impossible. For $r\gg r_{\rm S}$ this factor can be approximated as $h \approx 1$ and the gravitational term simplifies to
\begin{eqnarray}
\label{gravity}
{\bf f}_{G}&=&- \frac{\gamma^2 \xi \rho_0 G M_{*}}{r^2} {\bf \hat{e}}_r = -\frac{p_{0}^{3/4}\gamma^{5}v^{3}GM_{*}}{Q^{3/4}r^{5}}\Big(1+\frac{4 p_{0}^{1/4}Q^{3/4}\gamma v}{c^{2}r}\Big){\bf \hat{e}}_r\,.
\end{eqnarray}
It consists of two terms, of which the first one is proportional to $r^{-5}$ and the second one is proportional to $r^{-6}$. This combination again does not permit self-similar solutions, as all the other terms appearing in the momentum equation are proportional to $r^{-5}$. A possible way of taking partially into account gravity is by assuming a plasma at distances $r\gg r_{\rm S}$, with non-relativistic temperatures ($p\ll \rho_0 c^2$) and setting $\xi \approx 1$, so that in the inertia term the derivative of $\xi$ will be taken into account, but in the gravitational term it will be set to $\xi=1$. This allows separation of variables, but also adds an extra constraint on $Q$. In our solutions we decided not take into account gravity. This is not an absurd assumption, as we are interested in late stages of expanding systems where the fluid has reached relativistic velocities and has already expanded a lot so that it is not close enough to the central mass for gravity to have an important effect. 
We remark that while preparing this paper for publication, a similar study by \cite{TAM2009} appeared. These authors also attempted to include gravity; in fact the gravitational term is very important in their approach, since its magnitude is such as to balance the other forces for a {\emph{given}} poloidal magnetic field. However, they omitted a factor $\gamma \xi$ in the gravitational term of the momentum equation, compare our equation~(\ref{gravity}) with the last term of their equation~(2).
Our approach is different: we {\emph{find}} the poloidal magnetic field that corresponds to flows in which gravity is unimportant.

We now substitute the force densities found in equations~(\ref{inertia_term}) -- (\ref{em_term}) in the momentum equation to find  
\begin{eqnarray}
{\cal F}\nabla P + 16 \pi^3 \sin^2\theta \gamma^4 v^4  \nabla p_0 = 0 \,,
\label{mom1}
\end{eqnarray}
A direct consequence is that $\nabla p_0 \parallel \nabla P$, or,
\begin{eqnarray}
p_0=p_0(P) \,.
\end{eqnarray}
Equation~(\ref{mom1}) then becomes
(after substituting ${\cal F}$ from~\ref{GS})
\begin{eqnarray}
v^2 \frac{\partial^{2} P}{\partial v^{2}}
-\frac{2v^3}{1-v^2}\frac{\partial P}{\partial v}
+\frac{\sin \theta}{1-v^2} \frac{\partial}{\partial \theta} \left(\frac{1}{\sin \theta} \frac{\partial P}{\partial \theta} \right)  +\frac{v^2}{(1-v^2)^2} \beta \frac{{\rm d} \beta}{{\rm d} P}
+ 16 \pi^3 \sin^2\theta \frac{v^4}{(1-v^2)^3} \frac{{\rm d} p_0}{{\rm d}P}=0 \,.
\label{MOM2}
\end{eqnarray}
This is the necessary condition that the $r$ and $\theta$ components of the momentum equation (\ref{momentum}) are both zero, the $\phi$ component of the momentum equation is zero as shown by~(\ref{Teq0}). All pressure and inertia effects are included through the last term of the previous equation, which we shall call pressure-inertia term. In the case $p_0= $ const the ${\bf f}_{\rm I}$ and ${\bf f}_{p}$ forces cancel each other, and we are back in the force-free case ${\cal F}=0$.

As explained in Appendix~\ref{appA}, the only nontrivial semi-analytic solution
of the previous equation corresponds to
\begin{eqnarray}
P=g(v) \sin^2 \theta \,, 
\quad
\beta \frac{{\rm d} \beta}{{\rm d}P}= c_0 P \,, 
\quad
\frac{{\rm d} p_0}{{\rm d}P} = \frac{c_1}{16 \pi^3} \,,
\label{ss}
\end{eqnarray}
where $c_0$ and $c_1$ are constants.

Equations~(\ref{magnetic}), (\ref{electric}), (\ref{pressure}) and (\ref{density}) give the expressions of the physical quantities for the self-similar solution
\begin{eqnarray}
{\bf B}=
\frac{g}{\pi r^2} \cos \theta {\bf \hat{e}}_r 
-\frac{v g'}{2 \pi r^2} \sin \theta {{\bf \hat{e}_\theta}}
+\frac{g}{2 \pi r^2} \frac{\beta}{P} \gamma^2 v \sin \theta {{\bf \hat{e}_\phi}} \,,
\label{newB}
\end{eqnarray}
\begin{eqnarray}
{\bf E}=
\frac{g}{2\pi r^2} \frac{\beta}{P} \gamma^2 v^2 \sin \theta {{\bf \hat{e}_\theta}}
+\frac{v^2 g'}{2 \pi r^2} \sin \theta {{\bf \hat{e}_\phi}} \,,
\end{eqnarray}
the density and pressure are given by equations~(\ref{pressure}) and (\ref{density}) and we can substitute for $\beta$ and $p_{0}$
\begin{eqnarray}
\beta = \pm \left(c_0 P^2 + \beta_{00}\right)^{1/2} 
\,, \quad 
p_0=p_{00}+\frac{c_1}{16 \pi^3} g \sin^2 \theta 
\,,
\label{bp}
\end{eqnarray}
and a prime denotes derivative with respect to $v$.
Here $\beta_{00}$ and $p_{00}$ are constants, and $Q$ is a free function of $v$ and $\theta$.
The $\beta_{00}$ should vanish so that the azimuthal component of the magnetic field remains finite on the axis. Thus, $\beta = \pm c_0^{1/2} g \sin^2 \theta $ and $c_0$ must not be negative.
On the other hand $p_{00}$ should be a non-negative constant so that the pressure does not vanish on the axis. The positive (negative) sign of $c_3$ corresponds to increasing (decreasing) pressure as we move away from the axis, respectively. In addition, the sign of $c_1$ controls the sign of ${\cal F}$: they are opposite since equation~(\ref{mom1}) yields ${\cal F}+ c_1 \sin^2 \theta \gamma^4 v^4=0$. For $c_1>0$ the electromagnetic force points away from the axis and the sum of inertia and pressure forces points toward the axis; for $c_1<0$ the opposite.
Note that the freedom of an additive constant in the expression of the pressure means that the physical behaviour of the system does not change if we assume a background pressure which is proportional to $\gamma^{4}/(ct)^{4}$. The reason is that this part of the pressure introduces additional terms in the inertia and pressure gradient forces which cancel each other, see equations~(\ref{inertia_term})~and~(\ref{pressure_term}).

Then by substituting (\ref{ss}) into~(\ref{MOM2}), the latter becomes the following ordinary differential equation for $g(v)$ 
\begin{eqnarray}
v^2 g'' -\frac{2v^3 g'}{1-v^2}-\frac{2g}{1-v^2}
+\frac{c_0v^2 g}{(1-v^2)^2}
+\frac{c_1 v^{4}}{(1-v^{2})^3}=0\,,
\label{ode}
\end{eqnarray}
where the first three terms depend on the poloidal magnetic flux, the fourth is related to the toroidal magnetic field, and the fifth comes from the sum of inertia and pressure terms of the momentum equation.

Equation~(\ref{ode}) can be solved numerically for various values of the parameters $c_0$ and $c_1$. These parameters are related to the relative importance of the toroidal field and the fluid pressure and inertia compared with the poloidal field, see equations~(\ref{ss}). We present the results of the numerical solution in the next section. 

\section{Results}

We have solved numerically equation~(\ref{ode}), which is a second order ordinary differential equation. Our motivation is to describe a physical system where some magnetic flux emerges from a surface $v=v_{0}$ and expands uniformly within a sphere in the velocity space extending to $v_{\rm max}\approx 1$. We normalize $g(v)$ to the dimensionless function $\tilde{g}=g/g(v_{0})$, therefore the first boundary condition is $\tilde{g}(v_{0})=1$. This choice of normalization affects $c_{1}$ which is now substituted by the dimensionless $\tilde{c}_{1}=c_{1}/g(v_{0})$. The other boundary condition comes from the fact that the flux does not go further than $v_{\rm max}$, thus it is $\tilde{g}(v_{\rm max})=0$. Subject to these boundary conditions we are going to solve the equation for various combinations of the parameters appearing. However, before moving to these numerical solutions we investigate its asymptotic behaviour for $v\ll 1$.

\subsection{Asymptotic solutions for $v\ll 1$}

When we focus to the non-relativistic limit $v\ll 1$, the differential equation simplifies. The factor $\gamma=(1-v^{2})^{-1/2}$ is close to unity, thus the equation initially reduces to
\begin{eqnarray}
v^{2}\tilde{g}''-2v^{3}\tilde{g}'+(c_{0}v^{2}-2)\tilde{g}+\tilde{c}_{1}v^{4}=0.
\label{asym}
\end{eqnarray}
The Frobenius expansion at small $v$ with the first term $\tilde{g} \propto v^F$ gives indicial equation $(F+1)(F-2)=0$ for $F<4$.\footnote{There is also the possibility $F=4$, with the solution $\tilde{g}\approx -(c_1/10)v^4$. However, this case corresponds to overcollimated field lines and for that reason is rejected.} The solution with $F=2$ corresponds to cylindrical field lines, while the $F=-1$ to a dipolar magnetic field. Since we are interested in a physical system where the flux is generated by a  central source and then expands we study the dipolar solution with $F=-1$.

\subsection{Parameter study}

\begin{figure}
  \centering
    \includegraphics[width=0.5\textwidth]{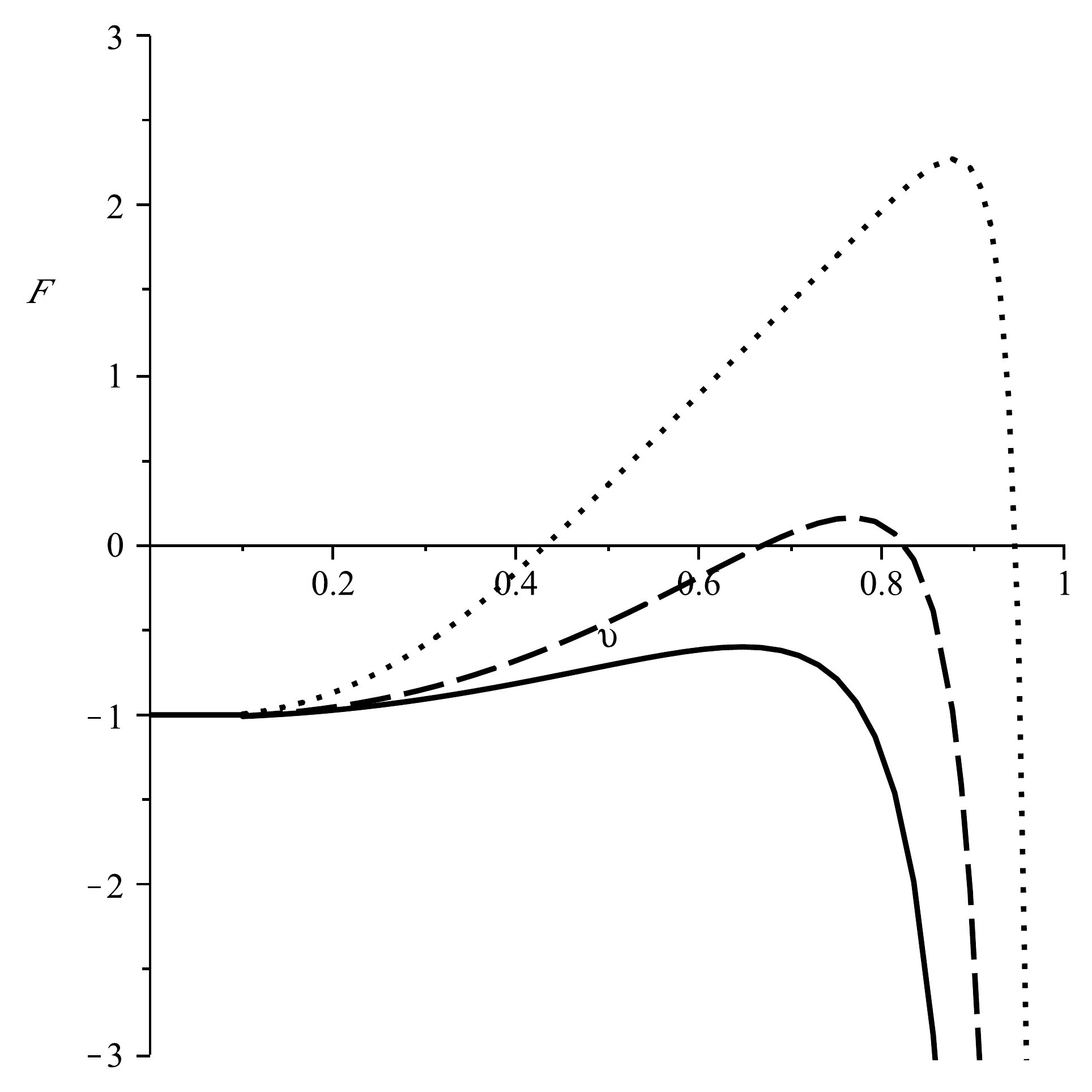}
     \caption{The expansion factor $F={\rm d} \ln \tilde{g}/{\rm d} \ln v$ according to the numerical solution of equation~(\ref{ode}) for $c_{0}=1$ and $\tilde{c}_{1}=0.1$. The solid line is for  $v_{\rm max}=0.95$, the dashed one for $v_{\rm max}=0.97$ and the dotted one for $v_{\rm max}=0.99$. All solutions converge to $-1$ for $v$ small so they have a dipolar behaviour. When $v \approx 1$ the expansion factor becomes very small and negative, so that the field lines close within the light sphere. In intermediate distances the ones for $v_{\rm max}=0.97$ and $v_{\rm max}=0.99$ have a positive $F$, so the field lines have x-type points and then focal points.}
\label{fig:g_c0_1}
\end{figure}

The relative importance of the physical quantities determining the behaviour of the fluid is parametrized by the two constants appearing on~(\ref{ode}). The ratio of $\beta$ over $P$ is $\pm c_0^{1/2}$, therefore a larger value of $c_{0}$ corresponds to a stronger toroidal component of the field. The definition of $\tilde{c}_{1}$, similarly, relates the plasma pressure-inertia to the magnetic field. The asymptotic behaviour of~(\ref{ode}) demonstrates that these terms are not important if $v$ is small and it is the poloidal flux emerging from the central dipole that dominates. As these terms are multiplied by powers of $\gamma$ the solution will depend strongly on both the choice of the parameters and on the choice of $v_{\rm max}$. The pressure-inertia term is multiplied by $\gamma^{6}$, thus at very high velocities it is this term that determines the solution, whereas in intermediate velocities it is the combination of the parameters chosen. Therefore the solution depends on three parameters $c_{0}$, $\tilde{c}_{1}$ and $v_{\rm max}$. In our parameter study we examine the problem for various combinations of the parameters, the cases studied are for $c_{0}=0, 0.5, 1$, for $\tilde{c}_{1}=0, 0.01, 0.1$ and for $v_{\rm max}=0.95, 0.97, 0.99$. We have plotted the expansion factor $F={\rm d} \ln \tilde{g}/{\rm d} \ln v$ for $\tilde{c}_{1}=0.1$ and $c_{0}=1$ (figure~\ref{fig:g_c0_1}). The results are similar for $c_{0}=0, 0.5$. The value of $F$ demonstrates the behaviour of the field lines; $F=-1$ corresponds to a dipole, and $F=2$ to a cylindrical field, $F=0$ with $F'>0$ gives the x-type points and $F=0$ with $F'<0$ gives the focal points of the field lines. 

We have chosen relatively small values for $c_{0}$ and $\tilde{c}_{1}$ so that $v_{\rm max}$ can be close to unity. Had we chosen larger values for these parameters, the flux would have been zero before reaching relativistic velocities. If we continue the integration further than $v_{\rm max}$, even for small values of $c_{0}$ and $\tilde{c}_{1}$, then $\tilde{g}$ will oscillate around zero, these oscillations correspond to closed loops causally disconnected from the base \citep{TS2007}.

\begin{figure}
  \centering
    \includegraphics[width=0.5\textwidth]{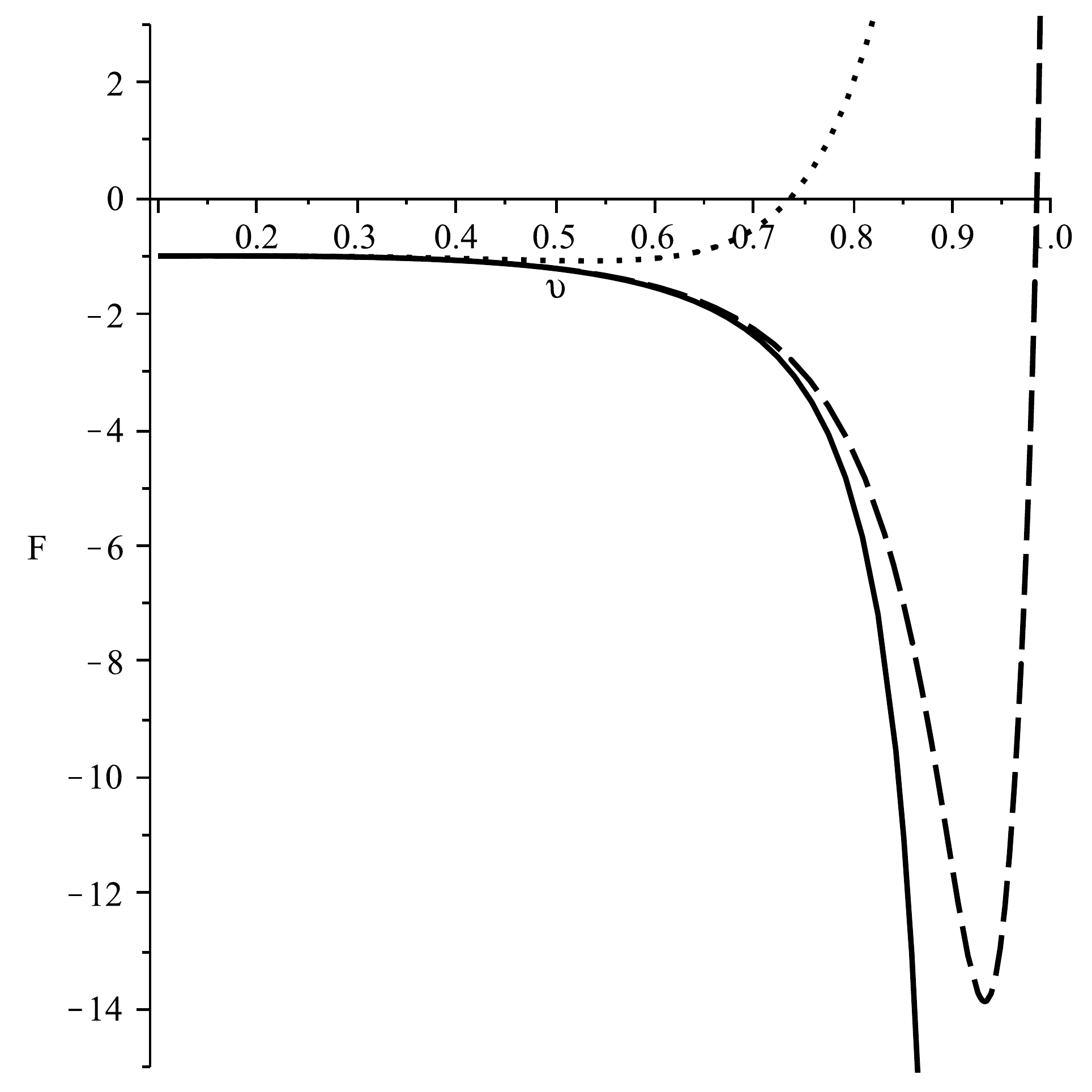}
     \caption{The expansion factor $F={\rm d} \ln \tilde{g}/{\rm d} \ln v$ according to the numerical solution of equation~(\ref{ode}) for negative $\tilde{c}_{1}$. Depending on the choice of the parameters the solution either diverges to infinity or becomes zero for $v<1$. The parameters chosen are $c_{0}=0.5$, $\tilde{g}(0.1)=1$ and $\tilde{g}'(0.1)=-9.97$ which are the same for all curves. Then the equation was integrated for $\tilde{c}_{1}=-0.01$ (solid line), $\tilde{c}_{1}=-0.1$ (dashed line) and $\tilde{c}_{1}=-0.5$ (dotted line). It is evident that when the pressure dominates over the other forces it requires more flux to achieve equilibrium.}
\label{fig:g_c1<0}
\end{figure}

We also study the problem for negative values of $\tilde{c}_{1}$, (figure~\ref{fig:g_c1<0}) this corresponds to systems where the pressure decreases as we move away from the axis. We find from~(\ref{bp}) that it is essential to include a background pressure $p_{00}$, so that none of the gas pressure or the density are negative.

\begin{figure}
  \centering
    \includegraphics[width=0.8\textwidth]{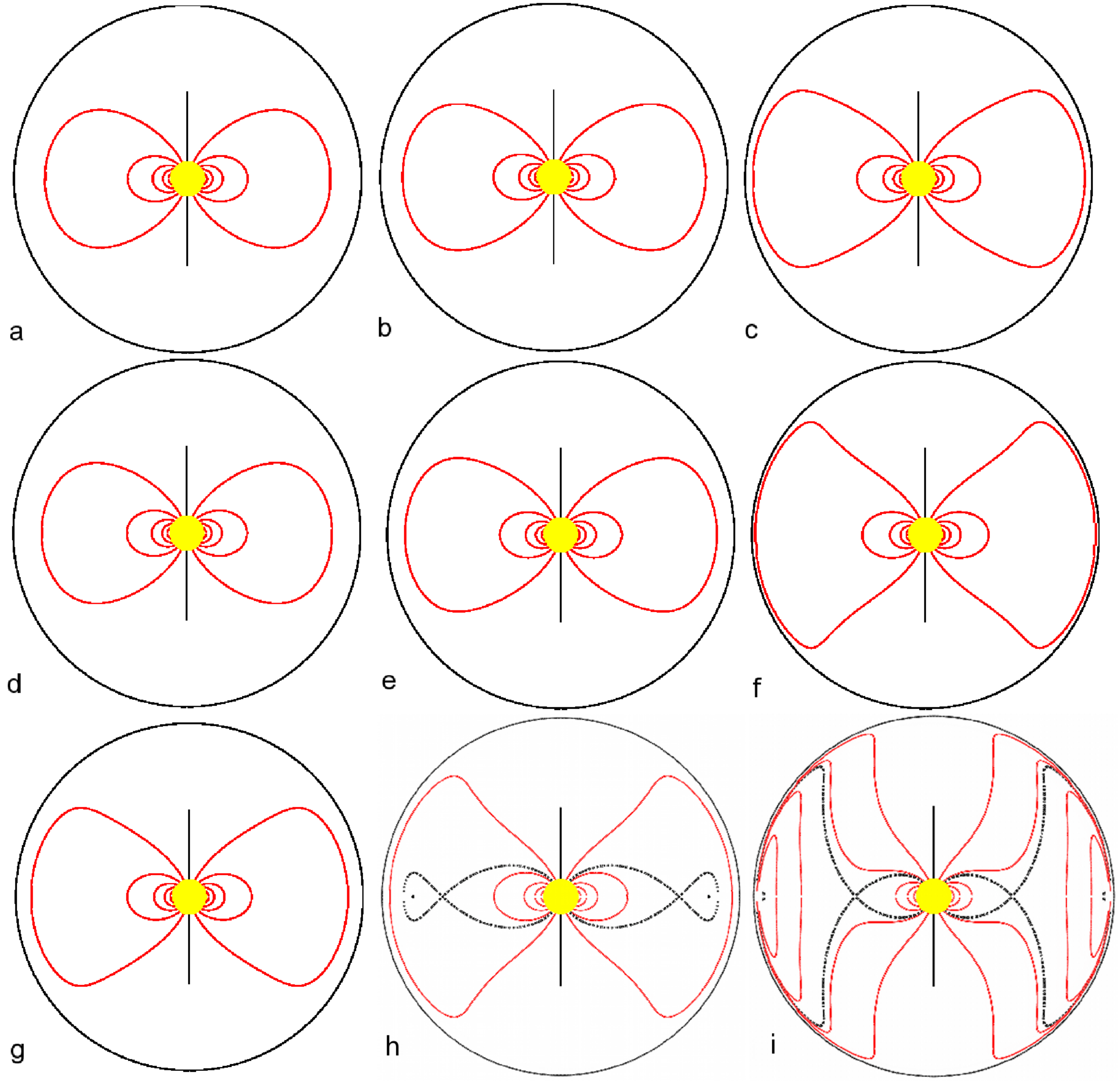}
     \caption{Plot of the poloidal field lines for $c_{0}=1.0$, and combinations of $\tilde{c}_{1}$ and $v_{\rm max}$. The first row (a, b, c) corresponds to $c_{1}=0$, the second (d, e, f) to $c_{1}=0.01$ and the third (g, h, i) to $c_{1}=0.1$. The first column (a, d, g) corresponds to $v_{\rm max}=0.95$, the second (b, e, h) to $v_{\rm max}=0.97$ and the third (c, f, i) to $v_{\rm max}=0.99$. The field lines near the origin have a dipole structure, but as they approach $v=v_{\rm max}$ they close, whereas in an ideal dipole there would not be such a boundary. The external circle marks the light sphere where the expansion velocity formally reaches the speed of light. The dotted line in (h) and (i) is the separatrix which corresponds to the local minimum of the flux and encircles the closed lobes appearing, the lobes have a central focus which lies at the equatorial plane.}
\label{fig:P_c0_1}
\end{figure}

The numerical solution for any combination of parameters verifies the fact that the field behaves like a dipole near the origin. Then, when it approaches the upper limit $v_{\rm max}$ the field deviates from the dipolar structure. When $v_{\rm max}$ comes closer to unity or $\tilde{c}_{1}$ is relatively large and positive there is more flux generated which forms closed loops. These loops are contained within a separatrix surface corresponding to the root of the expansion factor $F$ with $F'>0$; and have focal points which form a circle on the equatorial plane whose radius is determined by the position of the root of $F$ with $F'<0$. The physical explanation for the formation of these loops is that there is more electromagnetic pressure needed to force the flux to reach a higher velocity. A higher velocity leads to a greater Lorentz factor multiplying the rest mass of the plasma. Thus a small increase in the $v_{\rm max}$ leads to a dramatic increase in the pressure-inertia term and since the flux emerging from the spherical surface is limited, more flux has to be generated somehow in order to balance the forces. This extra flux appears in the form of these closed loops. The pressure-inertia term has also an effect on the collimation of the field: as it becomes more important the field lines become parallel to the axis, leading to a collimated magnetic field. The collimation and the closed loops are evident in figure~\ref{fig:P_c0_1} where the poloidal magnetic field lines (sections of surfaces of constant flux with a meridional plane) are plotted. There is also a very strong field near the $v_{\rm max}$. In this area the field has a very weak radial component whereas its $\theta$ component is large marking the turn over of the field lines, as they are not permitted to exceed $v_{\rm max}$. By comparison to the inertia-free case we have found that the addition of inertia has a more important effect, as we expected, because the pressure-inertia forces, parametrized by $\tilde{c}_{1}$ are multiplied in~(\ref{ode}) by a factor of $\gamma^{6}$, thus they are very sensitive to $v_{\rm max}$.  

\begin{figure}
  \centering
    \includegraphics[width=0.5\textwidth]{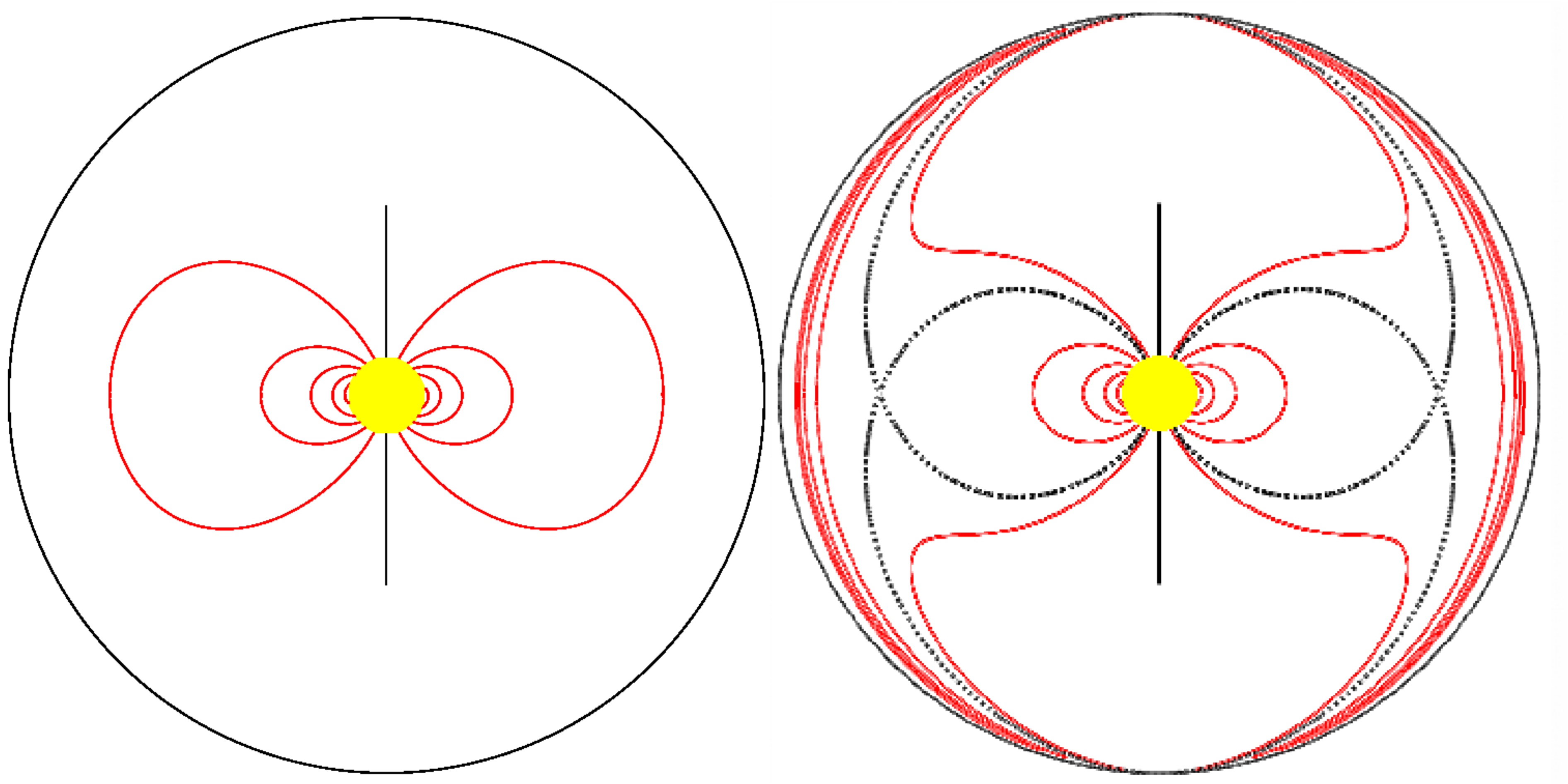}
     \caption{Left: Plot of the poloidal field lines for $c_{0}=0.5$, $\tilde{c}_{1}=-0.01$ and $\tilde{g}'(v_{0})=-9.97$. The pressure-inertia force is weak and causes little modification to the solution compared to the previous ones. Right: Plot of the poloidal field lines for $c_{0}=0.5$, $\tilde{c}_{1}=-0.5$ and $\tilde{g}'(v_{0})=-9.97$. The separatrix field line that corresponds to expansion factor $F=0$ is the dotted line. The field lines enclosed by the separatrix have the usual dipolar structure, however the ones that are not enclosed emerge from a monopole at $v=1$ and $\theta=-\pi$ and finish at a second monopole at $v=1$ and $\theta=\pi$. Fields containing monopoles are unphysical and thus unacceptable. These monopoles appear because we are trying to construct a field that is impossible, as we request uniform expansion and the total force to be zero. Thus we need more magnetic flux to balance the very strong force due to the inertia-pressure term. Since we allow only limited flux to emerge from the central sphere the extra flux emerges from the poles of the light sphere. This is the reason $\tilde{g}$ goes to infinity at $v=1$, depicting the need for extra flux.}
\label{fig:Nc1}
\end{figure}

When $\tilde{c}_{1}<0$ the system behaves differently. The direction of the pressure-inertia force term is opposite to the direction of the electromagnetic force arising from $\nabla P$, therefore it is their relative intensity that determines the fate of $\tilde{g}$. The details depend on the initial conditions and the parameters but there are two families of solutions. For relatively large $|\tilde{c}_{1}|$ the pressure-inertia term becomes strong compared to the electromagnetic terms of the momentum equation early enough then $\tilde{g}$ increases fast and diverges to infinity at $v=1$, this generates infinite pressure and flux, making this configuration unacceptable. On the other hand for relatively small $|\tilde{c}_{1}|$ the electromagnetic term dominates over the pressure-inertia term in~(\ref{ode}) and $\tilde{g}$ becomes zero for $v<1$, therefore the field is confined within a sphere and there are no infinite fields or pressure, figure~\ref{fig:Nc1}. These fields are acceptable.

\begin{figure}
  \centering
    \includegraphics[width=0.5\textwidth]{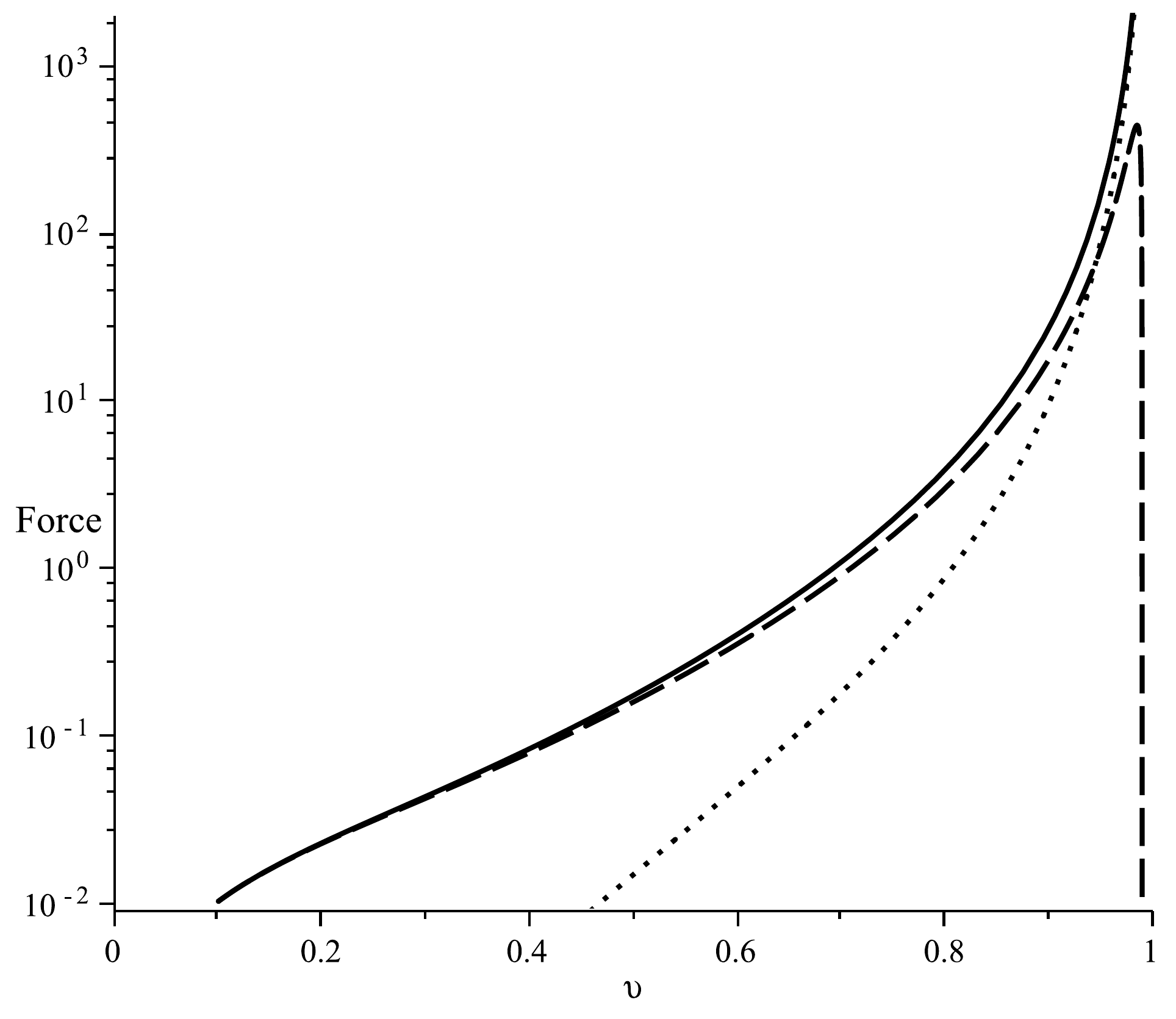}
     \caption{The terms of equation~(\ref{ode}) for $c_{0}=1$, $\tilde{c}_{1}=0.1$, $\tilde{g}(0.1)=1$ and $v_{\rm max}=0.99$. The solid line corresponds to the $|v^2 g'' -[2v^3/(1-v^2)]g'-[2/(1-v^20)] g|$ the absolute value of the force due to the poloidal field, the dashed line is force due to the toroidal field $c_0v^2/(1-v^2)^2g$ and the dotted line is $c_1 v^{4}/(1-v^{2})^3$. When $v$ is small the forces of the toroidal field balance the poloidal, however near the top it is the the forces due to the pressure that become important.}
\label{fig:Forces}
\end{figure}

In non-relavistic force-free MHD the field lines coincide with the current lines, whereas now with nonzero displacement current the picture is not so simple as there are also forces between the charge and the electric field, forces due to the gradient of pressure and inertia forces. However, in our uniformly expanding model the electromagnetic force is normal to the lines of constant $P$, since it is proportional to $\nabla P$, and as $\nabla P \parallel \nabla p_{0}$ so do the pressure-inertia forces. The relative importance of the various forces appears in the momentum equation~(\ref{ode}), in particular $v^2 g'' -[2v^3/(1-v^2)]g'-[2/(1-v^2)] g$ expresses the electromagnetic forces due to the poloidal magnetic field, $[c_0v^2/(1-v^2)^2]g$ is the electromagnetic forces due to the toroidal magnetic field, and $c_1 v^{4}(1-v^{2})^3$ are the forces due to pressure and inertia. We plot these terms for the case of $c_{0}=1$, $\tilde{c}_{1}=0.1$ and $v_{\rm max}=0.99$ in figure~\ref{fig:Forces}. The results for the relative intensity of forces are similar for other combinations of the parameters.  

\section{Physical quantities}

In this section we study the relation of the parameters appearing in the equations to physical quantities. The physical quantities we are interested in are the energy and the twist of the field lines. 

\subsection{Energy}

The magnetised fluid contains energy in four forms electromagnetic, kinetic, rest mass and thermal energy. The energy equation written in conservative form is $\partial T^{00} / \partial t + \nabla \cdot \left(c T^{0j} \hat x_j \right) = 0$ where $T^{\mu \nu}$ is the energy momentum tensor, (see e.g., \citealp{VK03a}), which gives
\begin{eqnarray}
&& \frac{\partial}{\partial t}\Big(\xi \gamma^2 \rho_0 c^2 - p+\frac{B^2+E^2}{8\pi}\Big)+ \nabla \cdot \Big(\xi\rho_0c^2 \gamma^2 c{\bf v} +\frac{c}{4\pi} {\bf E} \times {\bf B} \Big) =0 \,.
\label{energy_eq}
\end{eqnarray}
The first two terms inside the time derivative of the above equation represent the energy density of the fluid. It consists of the rest $\gamma\rho_0 c^2$, kinetic $\left(\gamma-1\right)\gamma\rho_0 c^2$, and thermal $\left(4\gamma^2-1\right)p$ energy densities. The last term inside the time derivative represents the energy density of the electromagnetic field. 
Similarly, the terms inside the space derivative correspond to the fluid energy flux and the Poynting flux. We now apply the Poynting theorem
\begin{eqnarray}
\frac{\partial}{\partial t}\Big(\frac{B^2+E^2}{8\pi}\Big)+\nabla \cdot \Big(\frac{c}{4\pi} {\bf E} \times {\bf B} \Big) =-\bm j \cdot {\bm E} \,,
\label{poynting_theorem}
\end{eqnarray}
equation~(\ref{energy_eq}) yields
\begin{eqnarray}
\frac{\partial}{\partial t}(\xi \gamma^2 \rho_0 c^2 - p)+\nabla \cdot (\xi\rho_0c^2 \gamma^2 c{\bf v}) ={\bf j} \cdot {\bf E} \,.
\end{eqnarray}
Thus the term ${\bf j} \cdot {\bf E}$ measures the energy transfer between the fluid and the electromagnetic field. In our model this term equals
\begin{eqnarray}
{\bf j} \cdot {\bf E} = {\bf j} \cdot ({\bf B} \times {\bf v}) = 
( {\bf j} \times {\bf B} ) \cdot {\bf v} = c {\bf v} \cdot {\bf f}_{\rm em}
=\frac{c \gamma^4 v^6 g' \sin^2 \theta}{r^5} \frac{{\rm d}p_0}{{\rm d}P}
=\frac{c_1 c \gamma^4 v^6 g' \sin^2 \theta}{16 \pi^3 r^5}  \,.
\end{eqnarray}
The sign of this quantity shows the flow of the energy, when it is positive energy flows from the field to the fluid and vice versa. This sign depends only the product $c_{1}g'$, since all the other terms are positive.
\begin{figure}
  \centering
    \includegraphics[width=0.8\textwidth]{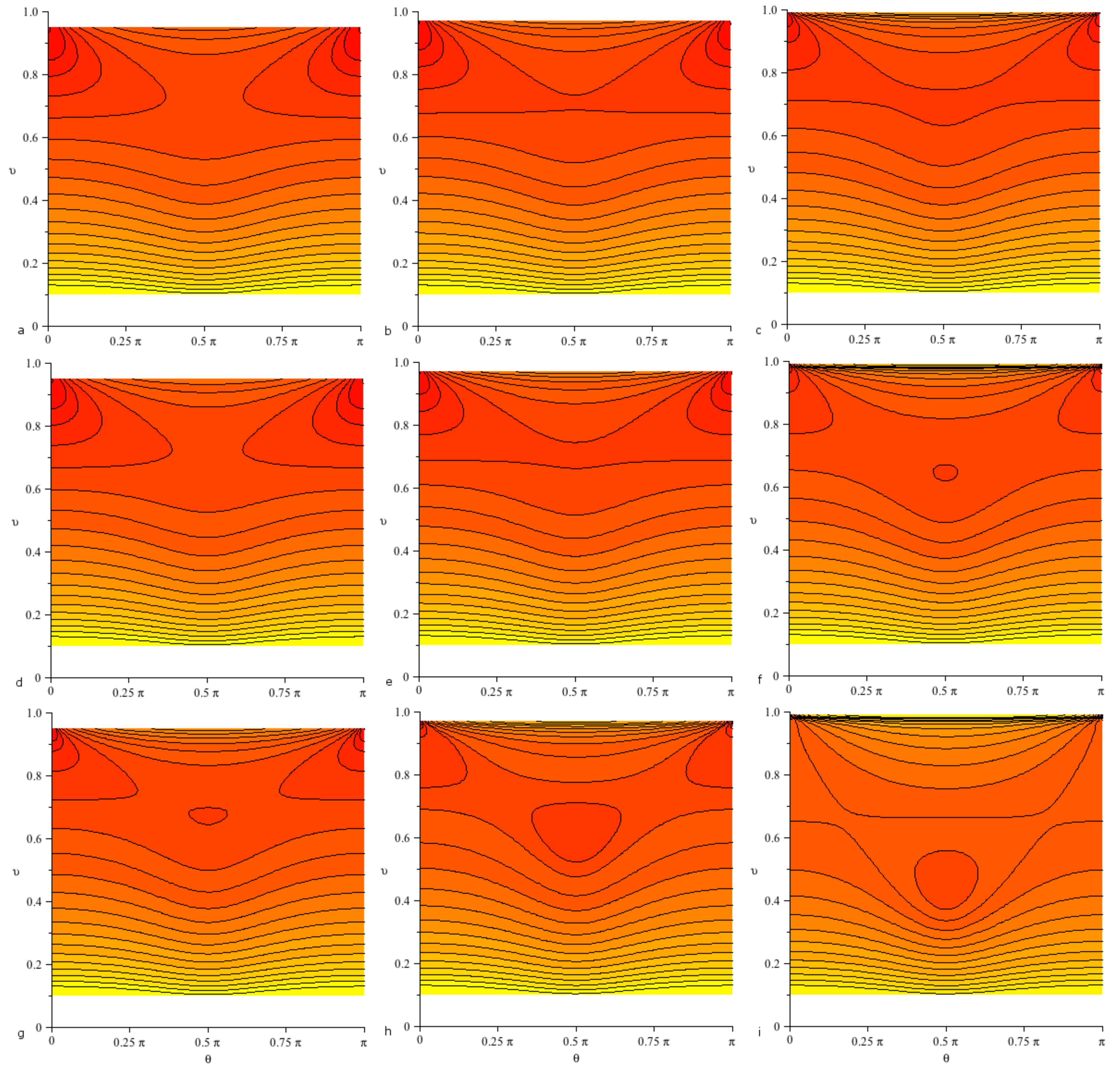}
     \caption{Plot of the energy density of the electromagnetic field for $c_{0}=1$. The first line (a, b, c) has $c_{1}=0$, the second (d, e, f) $c_{1}=0.01$ and the third (g, h, i) $c_{1}=1$. The first column (a, d, g) reaches $v_{\rm max}=0.95$, the second column (b, e, h) $v_{\rm max}=0.97$ and the third (c, f, i) $v_{\rm max}=0.99$. }
\label{fig:e_c0_1}
\end{figure}
The energy density of the electromagnetic field is plotted in figure~\ref{fig:e_c0_1}, the energy density of the fluid depends on the choice of $p_{00}$ and although it is related to the energy density of the electromagnetic field it cannot be defined in an rigorous way. 

\subsection{Twist}

We now have an expression for the fields everywhere in space. We can evaluate the twist of the magnetic field lines. The lines of force of the magnetic field are determined by the relation:
\begin{eqnarray}
\frac{{\rm d}r}{B_{r}}=\frac{r {\rm d} \theta}{B_{\theta}}=\frac{r \sin \theta {\rm d} \phi}{B_{\phi}}\,.
\label{lines}
\end{eqnarray}
We substitute the expressions for the magnetic field~(\ref{newB}) in the equation for the field lines~(\ref{lines}). From the first equality we take $g\sin^{2} \theta=P_{i}$, where the constant $P_{i}$ is the value of $P$ for this particular field line. Then we equate the first and the third part of equation~(\ref{lines})
\begin{eqnarray}
{\rm d} \phi =\pm \frac{\gamma^{2}c_{0}^{1/2}{\rm d}r}{2ct(1-P_{i}/g)^{1/2}}\,.
\label{lines2}
\end{eqnarray}
In this expression (\ref{lines2}) $\gamma$ and $g$ depend on $v$ whereas the integral is to be done in $r$, however we can change the variable of integration from $r$ to $v$, by including the $ct$ of the denominator in the differential. The limits of integration are $v=v_{0}$, the surface the field line emerges from, and the maximum distance $v=v_{i}$ it reaches in velocity space. This upper limit of integration is the solution of the equation $g(v_{i})=P_{i}$. The twist for a given field line between two points in velocity space is constant with time as it only depends on $v$. The physical meaning of this result is that the field lines have already been twisted before the expansion starts and then they merely expand radially. By symmetry the total twist for the outgoing part (from $v_{0}$ to $v_{i}$) and the incoming part (from $v_{i}$ to $v_{0}$) is twice the integral from $v_{0}$ to $v_{i}$. The total twist is
\begin{eqnarray}
\Phi=\pm c_{0}^{1/2}\int_{v_{0}}^{v_{i}}\frac{\gamma^{2} {\rm d}v}{(1-P_{i}/g)^{1/2}}\,.
\label{PHI}
\end{eqnarray}

The twist explicitly depends on $c_{0}$ as this determines the ratio of the toroidal to the poloidal field; in addition, the form of $g$ that appears in the integral depends on all the parameters and the boundary conditions, so apart from $c_{0}$ the choice of both $c_{1}$ and $v_{\rm max}$ have an effect on the value of the twist. The presence of $P_{i}$ in the integral demonstrates that the twist depends also on the individual field line we choose to study. The field line that emerges from the pole corresponds to $P_{i}=0$ and the twist only depends on $c_{0}$ and $v_{\rm max}$, as the dependence on $g$ is annulled since it is multiplied by zero.

\section{Discussion}

In this paper we have studied the problem of the uniform expansion of a magnetised fluid. The configuration consists of a polytrope with $\Gamma=4/3$ corresponding to a completely degenerate electron gas in the extreme relativistic limit and an electromagnetic field that satisfies ideal MHD. This value is the only one allowing separation of variables, we remark that the same type of polytrope was chosen by \cite{L1982} and \cite{TS2007} who studied non-relativistic MHD flows. On this paper we study solutions that are connected to the origin, whereas \cite{TS2007} study fields that form magnetic islands disconnected from the origin. Their study is more appropriate for isolated magnetic plasmoids that have been twisted and lost causal connection with the parent object. 

The constraint of uniform expansion is introduced through the use of the dimensionless variable $v=r/ct$  which characterises each element of the magnetised fluid, so that it moves with constant velocity and suffers no acceleration; thus, the net force is zero everywhere but not the distinct forces due to the fluid pressure and inertia and the electromagnetic fields. In our model we take into account the electromagnetic interaction, the force due to the pressure gradient and the inertia force. These forces produce work so energy is exchanged between the electromagnetic field and the fluid. This combination of assumptions leads to separation of the partial differential equations of the problem and to semi-analytical solutions. We do not include gravity in this model. In the presence of gravity the equations do not separate unless we constrained our study in non-relativistic temperatures ($p \ll \rho c^{2}$) and distances $r \gg r_{S}$. 

This structure is powered by an increasing magnetic dipole at the origin which releases new flux, thus a small loop of current lies at $r=0$ and the current increases with time. A mechanism that may provide an increasing magnetic field is the Poynting Robertson battery, this mechanism has been proposed to operate in AGNs by \cite{CC2009}. This solution has the basic properties of a collimated jet, however in nature the exact form of the jet shall depend on the interaction with the external medium and the observational results on the radiation mechanisms which are not investigated in this paper. 

Comparing our results to the ones by \cite{TAM2009} we find that they are similar for small $v$, where the dipolar form of the solution dominates. However when $v$ becomes large they differ significantly. This is mainly of the different approach, we solve the problem making the assumption that the magnetic flux emerges from the origin and reaches a maximum expansion velocity and then we solve for the flux function, whereas they impose a poloidal field and then they solve for the other components of the equation.  

This physical system covers the late stage of the relativistic expansion of a magnetised fluid that has been already accelerated and each shell expands with constant velocity. As the velocity is proportional to the distance from the origin the inner shells do not overtake the outer ones, and there are no collisions. In this model the inlet boundary ($v_{0}$) is also moving, so this cannot be identified with a surface remaining still in real space, however $v_{0}$ can be chosen to be small and to move slowly compared to the  rest of the configuration. This study may be applied to systems that occur after the explosion of an object which contained a strong magnetic field.  

An interesting property of this model is that for some sets of parameters $g'$ becomes positive (leading to a positive expansion factor $F=vg'/g$) for intermediate velocities, meaning that the field is collimated there. At larger velocities the $g'$ becomes again negative so that the lines are closed inside the light sphere.

The demand for semi-analytical and separable solutions sets constraints in the range of physical configurations we can describe. If we are to study an accelerating or decelerating expansion it is essential to give up the uniform expansion parameter $v=r/ct$. In appendix B we show that the force equation does not admit self-similar solutions for an arbitrary combination of $r$ and $t$. 

In this paper we have found analytical solutions for a complicated relativistic MHD problem. We are aware that there is a gap to bridge between the observed radiation from a relativistically expanding magnetised fluid and our idealised model. We suggest that our work can be used a stepping stone for future studies and the check of the validity of MHD simulations. The results of the simulations can be compared with our solutions for consistency. We also remark that these structures allow the exchange of energy between the field and the fluid. This is a cooling/heating mechanism for the plasma and collimation/decollimation of the magnetic field respectively.

\section*{Acknowledgements} 
The authors are grateful to Professors Kanaris Tsinganos and Donald Lynden-Bell for illuminating discussions and insightful comments. KNG is grateful to the section of Astrophysics, Astronomy and Mechanics of the Department of Physics of the University of Athens, where he was a research visitor during autumn 2008 and part of this research was done. 

\begin{appendices}
\section{~Solutions of equation~(\ref{MOM2})}
\label{appA}

We seek separable solutions of equation~(\ref{MOM2}), of the self-similar form $P=P(\alpha)$, where $\alpha=g(v) \sin^2 \theta$.
By using $\alpha$ instead of $\theta$ we may transform from the pair of independent variables $(v\,,\theta)$ to the $(v\,,\alpha)$.
With the following elementary relations valid for any function $\Phi$,
\begin{eqnarray}
&&\frac{\partial \Phi(v\,,\theta)}{\partial v}= 
\frac{\partial \Phi(v\,,\alpha)}{\partial v} +
\frac{g'}{g} \alpha \frac{\partial \Phi(v\,,\alpha)}{\partial \alpha} \,,
\nonumber \\
&&\frac{\partial \Phi(v\,,\theta)}{\partial \theta}= 
2 \frac{\cos \theta}{\sin \theta}
\alpha \frac{\partial \Phi(v\,,\alpha)}{\partial \alpha} \,,
\end{eqnarray}
we may rewrite~(\ref{MOM2}), after dividing with $\alpha P'$, as 
\begin{eqnarray}
v^2 g'' -\frac{2v^3}{1-v^2}g' -\frac{2}{1-v^2}g =-\frac{v^2 g}{(1-v^2)^2} \frac{\beta}{\alpha P'} \frac{{\rm d} \beta}{{\rm d}P} -\frac{v^{4}}{(1-v^{2})^3} 16 \pi^3 \frac{1}{P'}\frac{{\rm d} p_0}{{\rm d}P}
\nonumber \\
-\frac{4g^2}{1-v^2} \frac{P''}{P'} +\left[\frac{4g}{1-v^2}-\frac{(vg')^2}{g}\right] \frac{\alpha P''}{P'}
\,,
\label{sum_products}
\end{eqnarray}
where primes denote derivative with respect to $v$ or $\alpha$. Note that the division with $\alpha P'$ is possible because this expression cannot be zero (this would mean that $P=$ const., or equivalently that the poloidal magnetic field vanishes).

We first examine two trivial cases.

The first trivial case corresponds to $g=\lambda = $ const. In this case~(\ref{sum_products}) yields $\beta=$ const, $p_0=$ const, and $P=C(1-\cos \theta)$, corresponding to a pure poloidal monopolar, force-free magnetic field.

A second trivial case corresponds to
\begin{eqnarray}
g=\lambda \frac{v^2}{1-v^2}\,,
\nonumber
\end{eqnarray}
where $\lambda=$ const. In this case~(\ref{sum_products}) yields
\begin{eqnarray} 
\frac{dp_0}{d\alpha}=\frac{-\lambda}{16 \pi^3}\left[
\frac{\beta}{\alpha} \frac{{\rm d}\beta}{{\rm d}\alpha}
+4 \left(P'\right)^2 + 4 \lambda P' P'' + 4 \alpha P' P''
\right]\,.
\nonumber
\end{eqnarray}
This solution corresponds to cylindrical poloidal magnetic field in the $v\ll 1$ regime, however this solution contains magnetic monopoles at $v=1$ and $\theta=0, \pi$ and it is unphysical.

In the general case, equation~(\ref{sum_products}) yields that there are constants $c_0$, $c_1$, $c_2$, $c_3$, such that (see Appendix~B in \citealp{VT97})
\begin{eqnarray}
v^2 g'' -\frac{2v^3}{1-v^2}g' -\frac{2}{1-v^2}g =
-\frac{c_0v^2 g}{(1-v^2)^2}
-\frac{c_1 v^{4}}{(1-v^{2})^3} 
-\frac{4 c_2 g^2}{1-v^2}
+c_3\left[\frac{4g}{1-v^2}-\frac{(vg')^2}{g}\right]
\,.
\label{veq1}
\end{eqnarray} 
Equation~(\ref{sum_products}) then becomes
\begin{eqnarray}
\left[\frac{(vg')^2}{g}-\frac{4g}{1-v^2}\right]
\left(\frac{\alpha P''}{P'} -c_3 \right) =
\frac{v^2 g}{(1-v^2)^2} 
\left(c_0 -\frac{\beta}{\alpha P'} \frac{{\rm d} \beta}{{\rm d}P} \right)
+\frac{v^{4}}{(1-v^{2})^3} 
\left(c_1- 16 \pi^3 \frac{1}{P'}\frac{{\rm d} p_0}{{\rm d}P} \right)\nonumber \\
+\frac{4g^2}{1-v^2} 
\left(c_2-\frac{P''}{P'} \right)
\,,
\label{sum_productsA}
\end{eqnarray}
i.e., it becomes a sum of four products of a function of $v$ with a 
function of $\alpha$.

We distinguish two possibilities. The first is that $\alpha P'' / P' -c_3 \neq 0$. In that case we can divide~(A.4) by $\alpha P'' / P' -c_3$, and find that there are constants $c_4$, $c_5$ and $c_6$ such that $(vg')^2/g-4g/(1-v^2)=c_4v^2 g/(1-v^2)^2 +c_5 v^{4}/(1-v^{2})^3 + 4 c_6 g^2/(1-v^2)$. However, it is highly unlikely that the last equation has solutions that satisfy also~(\ref{veq1}), except for the trivial cases $g=\lambda$ and $g=\lambda v^2/(1-v^2)$ considered above. So we are left with the second possibility, which is to have $\alpha P'' / P' -c_3 = 0$, with solution\footnote{The solution of $\alpha P'' / P' -c_3 = 0$ is that $P'$ is proportional to $\alpha^{c_3}$. However, without loss of generality we can assume that the constant of proportionality is unity.} $P'=\alpha^{c_3}$. Equation~(\ref{sum_productsA}) then becomes
\begin{eqnarray}
\left(g\frac{1-v^2}{v^2}\right)^2 \left(4\frac{c_3}{\alpha}-4c_2 \right)=
\left(g\frac{1-v^2}{v^2}\right)
\left(c_0 -\frac{\beta}{\alpha^{c_3+1}} \frac{{\rm d} \beta}{{\rm d}P} \right)+
\left(c_1- \frac{16 \pi^3}{\alpha^{c_3}}\frac{{\rm d} p_0}{{\rm d}P} \right)
\,.
\label{sum_productsB}
\end{eqnarray}

We again distinguish the following two possibilities. 
The first corresponds to the case where $c_2$ or $c_3$ is nonzero. In this case, by dividing~(\ref{sum_productsB}) by $4c_3/\alpha -4c_2$ we find that $[g(1-v^2)/v^2]^2$ equals a linear combination of $g(1-v^2)/v^2$ and a constant. This means that $g(1-v^2)/v^2 = $ const., a case that we already considered above. 
The second possibility is to have $c_2=c_3=0$, in which case $P=\alpha$. From equation~(\ref{sum_productsB}) we find $\beta {\rm d} \beta/{\rm d}P = c_0 \alpha$ and $16 \pi^3 {\rm d} p_0/{\rm d}P = c_1$, while~(\ref{veq1}) gives~(\ref{ode}).

In this appendix we have chosen that the angular part of the solution is proportional to $\sin^{2}\theta$, this corresponds to a dipole field. The differential operator $\sin \theta \partial/\partial \theta \left(\frac{1}{\sin \theta} \partial/\partial \theta \right)$ admits in general eigenfunctions in the form of $\sin\theta{\rm d}P_{l}(\cos \theta)/{\rm d} \theta$, where $P_{l}$ is the Legendre Polynomial of order $l$. The case studied above is for $l=1$. Assuming a linear form on $\beta(P) {\rm d}\beta/{\rm d}P= c_{0}P$ and $16\pi^{3} {\rm d}p_{0}/{\rm d}P=c_{1}$ we find that we are constrained only to the dipole solution. This is because~(\ref{MOM2}) reduces to the following form:
\begin{eqnarray}
\lambda(v)\sin\theta  \frac{{\rm d}P_{l}(\cos \theta)}{{\rm d} \theta} + c_{1} \sin^2\theta \frac{v^4}{(1-v^2)^3}=0 \,.
\end{eqnarray}

As the pressure-inertia term is multiplied by $\sin^{2} \theta$ the only acceptable solution is this of a dipole for $l=1$. In the absence of pressure and inertia there are acceptable solutions of higher order multipoles, see appendix B of \cite{GL2008} for more details. 

\section{~Solutions for arbitrary combination of $r$ and $t$}
\label{appB}
     
In this appendix we study whether it is possible to have separation of variables for a dimensionless parameter other than $v=r/(ct)$. Let us assume that we are looking for self-similar solutions of the form $v=r/R(t)$ where $R(t)$ is an arbitrary function of $t$ which has dimensions of length. Let us assume a magnetic field of the form of equation~(\ref{magnetic}). We shall follow step by step the process described in section 2, the electric field that satisfies the induction equation is ${\bf E}=-[\dot{R}/(cR)] {\bf \hat{e}}_{r} \times {\bf B}$. The gas pressure and inertia forces contribute only to the $r$ and the $\theta$ components of the momentum equation, whereas it is only the electromagnetic forces that contribute to the $\phi$ component of the momentum equation. We equate the $\phi$ component to zero, $(j^{0} {\bf E}+ {\bf j} \times {\bf B})\cdot {\bf \hat{e}_{\phi}}=0$. Unlike equation~(\ref{Teq0}) the equation we take for it is more complicated
\begin{eqnarray}
(v^{3}\dot{R}-v)\frac{\partial T}{\partial \theta}\frac{\partial P}{\partial v}+(v-v^{3}\dot{R}^{2}) \frac{\partial T}{\partial v} \frac{\partial P}{\partial \theta}+(v^{2}R\ddot{R}-v^{2}\dot{R}-1)T\frac{\partial P}{\partial \theta}=0\,,
\label{Rarbitrary}
\end{eqnarray}
where $\dot{R}={\rm d}R/{\rm d}t$. On division by $(v^{3}\dot{R}^{2}-v)$, the first and the second term of the above equation only depend on $v$ and $\theta$, therefore we expect the third term to have no dependence on $t$. This clearly happens when $R$ is linear function of $t$ and this is the condition we imposed in the first section of the paper, and indeed when substitute that in equation~(\ref{Rarbitrary})  we return to equation~(\ref{Teq0}).
\end{appendices}

\bibliographystyle{gGAF}

\bibliography{paper1}

\begin{thebibliography}{28}
\providecommand{\natexlab}[1]{#1}

\bibitem[\protect\citeauthoryear{{Blandford} and {Payne}}{1982}]{BP1982}
{Blandford}, R.D. and {Payne}, D.G., {Hydromagnetic flows from accretion discs
  and the production of radio jets}. {\itshape \mnras} 1982, \textbf{199},
  883--903.

\bibitem[\protect\citeauthoryear{{Chiueh} {\itshape{et~al.}}}{1991}]{CLB1991}
{Chiueh}, T., {Li}, Z. and {Begelman}, M.C., {Asymptotic structure of
  hydromagnetically driven relativistic winds}. {\itshape \apj} 1991,
  \textbf{377}, 462--466.

\bibitem[\protect\citeauthoryear{{Contopoulos}
  {\itshape{et~al.}}}{2009}]{CC2009}
{Contopoulos}, I., {Christodoulou}, D.M., {Kazanas}, D. and {Gabuzda}, D.C.,
  {The Invariant Twist of Magnetic Fields in the Relativistic Jets of Active
  Galactic Nuclei}. {\itshape \apjl} 2009, \textbf{702}, L148--L152.

\bibitem[\protect\citeauthoryear{{Contopoulos}}{1994}]{C1994}
{Contopoulos}, J., {Magnetically driven relativistic jets and winds: Exact
  solutions}. {\itshape \apj} 1994, \textbf{432}, 508--517.

\bibitem[\protect\citeauthoryear{{Contopoulos}}{1995}]{C1995}
{Contopoulos}, J., {Force-free Self-similar Magnetically Driven Relativistic
  Jets}. {\itshape \apj} 1995, \textbf{446}, 67--74.

\bibitem[\protect\citeauthoryear{{De Villiers}
  {\itshape{et~al.}}}{2005}]{D2005}
{De Villiers}, J., {Hawley}, J.F., {Krolik}, J.H. and {Hirose}, S.,
  {Magnetically Driven Accretion in the Kerr Metric. III. Unbound Outflows}.
  {\itshape \apj} 2005, \textbf{620}, 878--888.

\bibitem[\protect\citeauthoryear{{Fendt}}{1997}]{F1997}
{Fendt}, C., {Collimated jet magnetospheres around rotating black holes.
  General relativistic force-free 2D equilibrium.}. {\itshape \aap} 1997,
  \textbf{319}, 1025--1035.

\bibitem[\protect\citeauthoryear{{Fendt} and {Greiner}}{2001}]{FG2001}
{Fendt}, C. and {Greiner}, J., {Magnetically driven superluminal motion from
  rotating black holes. Solution of the magnetic wind equation in Kerr metric}.
  {\itshape \aap} 2001, \textbf{369}, 308--322.

\bibitem[\protect\citeauthoryear{{Gourgouliatos}}{2009}]{G2009}
{Gourgouliatos}, K.N., {Relativistically expanding cylindrical electromagnetic
  fields}. {\itshape \mnras} 2009, \textbf{396}, 2399--2404.

\bibitem[\protect\citeauthoryear{{Gourgouliatos} and
  {Lynden-Bell}}{2008}]{GL2008}
{Gourgouliatos}, K.N. and {Lynden-Bell}, D., {Fields from a relativistic
  magnetic explosion}. {\itshape \mnras} 2008, \textbf{391}, 268--282.

\bibitem[\protect\citeauthoryear{{Grad} and {Rubin}}{1958}]{GR1958}
{Grad}, H. and {Rubin}, H., {Hydromagnetic Equilibria and Force-Free Fields};
  in {\itshape Proceedings of the Second United Nations International
  Conference on the Peaceful Uses of Atomic Energy, Geneva}, Vol. ~31 1958, pp.
  190--197.

\bibitem[\protect\citeauthoryear{{Heyvaerts} and {Norman}}{2003}]{HN2003}
{Heyvaerts}, J. and {Norman}, C., {Global Asymptotic Solutions for Relativistic
  Magnetohydrodynamic Jets and Winds}. {\itshape \apj} 2003, \textbf{596},
  1240--1255.

\bibitem[\protect\citeauthoryear{{Li} {\itshape{et~al.}}}{1992}]{CLB1992}
{Li}, Z., {Chiueh}, T. and {Begelman}, M.C., {Electromagnetically driven
  relativistic jets - A class of self-similar solutions}. {\itshape \apj} 1992,
  \textbf{394}, 459--471.

\bibitem[\protect\citeauthoryear{{Low}}{1982}]{L1982}
{Low}, B.C., {Self-similar magnetohydrodynamics. I - The gamma = 4/3 polytrope
  and the coronal transient}. {\itshape \apj} 1982, \textbf{254}, 796--805.

\bibitem[\protect\citeauthoryear{{Lynden-Bell} and {Boily}}{1994}]{LB1994}
{Lynden-Bell}, D. and {Boily}, C., {Self-Similar Solutions up to Flashpoint in
  Highly Wound Magnetostatics}. {\itshape \mnras} 1994, \textbf{267}, 146--152.

\bibitem[\protect\citeauthoryear{{Meliani}
  {\itshape{et~al.}}}{2006}]{Meliani2006}
{Meliani}, Z., {Sauty}, C., {Vlahakis}, N., {Tsinganos}, K. and {Trussoni}, E.,
  {Nonradial and nonpolytropic astrophysical outflows. VIII. A GRMHD
  generalization for relativistic jets}. {\itshape \aap} 2006, \textbf{447},
  797--812.

\bibitem[\protect\citeauthoryear{{Mobarry} and {Lovelace}}{1986}]{ML1986}
{Mobarry}, C.M. and {Lovelace}, R.V.E., {Magnetohydrodynamic flows in
  Schwarzschild geometry}. {\itshape \apj} 1986, \textbf{309}, 455--466.

\bibitem[\protect\citeauthoryear{{Prendergast}}{2005}]{P2005}
{Prendergast}, K.H., {Relativistically expanding axisymmetric self-similar
  force-free fields}. {\itshape \mnras} 2005, \textbf{359}, 725--728.

\bibitem[\protect\citeauthoryear{{Sauty} and {Tsinganos}}{1994}]{ST1994}
{Sauty}, C. and {Tsinganos}, K., {Nonradial and nonpolytropic astrophysical
  outflows III. A criterion for the transition from jets to winds}. {\itshape
  \aap} 1994, \textbf{287}, 893--926.

\bibitem[\protect\citeauthoryear{{Sedov}}{1946}]{S1946}
{Sedov}, C.I., {Propagation of strong shock waves}. {\itshape Journal of
  Applied Mathematics and Mechanics} 1946, \textbf{10}, 241--259.

\bibitem[\protect\citeauthoryear{{Shafranov}}{1958}]{S1958}
{Shafranov}, V.D., {On Magnetohydrodynamical Equilibrium Configurations}.
  {\itshape Soviet Journal of Experimental and Theoretical Physics} 1958,
  \textbf{6}, 545--554.

\bibitem[\protect\citeauthoryear{{Shafranov}}{1966}]{S1966}
{Shafranov}, V.D., {Plasma Equilibrium in a Magnetic Field}. {\itshape Reviews
  of Plasma Physics} 1966, \textbf{2}, 103--151.

\bibitem[\protect\citeauthoryear{{Takahashi}
  {\itshape{et~al.}}}{2009}]{TAM2009}
{Takahashi}, H.R., {Asano}, E. and {Matsumoto}, R., {Relativistic expansion of
  magnetic loops at the self-similar stage}. {\itshape \mnras} 2009,
  \textbf{394}, 547--568.

\bibitem[\protect\citeauthoryear{{Taylor}}{1950}]{T1950}
{Taylor}, G., {The Formation of a Blast Wave by a Very Intense Explosion. I.
  Theoretical Discussion}. {\itshape Royal Society of London Proceedings Series
  A} 1950, \textbf{201}, 159--174.

\bibitem[\protect\citeauthoryear{{Tsui} and {Serbeto}}{2007}]{TS2007}
{Tsui}, K.H. and {Serbeto}, A., {Time-dependent Magnetohydrodynamic
  Self-similar Extragalactic Jets}. {\itshape \apj} 2007, \textbf{658},
  794--803.

\bibitem[\protect\citeauthoryear{{Vlahakis} and {K{\"o}nigl}}{2003}]{VK03a}
{Vlahakis}, N. and {K{\"o}nigl}, A., {Relativistic Magnetohydrodynamics with
  Application to Gamma-Ray Burst Outflows. I. Theory and Semianalytic
  Trans-Alfv{\'e}nic Solutions}. {\itshape \apj} 2003, \textbf{596},
  1080--1103.

\bibitem[\protect\citeauthoryear{{Vlahakis} and {Tsinganos}}{1997}]{VT97}
{Vlahakis}, N. and {Tsinganos}, K., {On the topological stability of
  astrophysical jets}. {\itshape \mnras} 1997, \textbf{292}, 591--600.

\bibitem[\protect\citeauthoryear{{Vlahakis} and {Tsinganos}}{1998}]{VT1998}
{Vlahakis}, N. and {Tsinganos}, K., {Systematic construction of exact
  magnetohydrodynamic models for astrophysical winds and jets}. {\itshape
  \mnras} 1998, \textbf{298}, 777--789.

\end{thebibliography}

\label{lastpage}

\end{document}